\documentclass[amsmath,amssymb,goupaddress,prb,aps,showpacs,showkeys,twocolumn]{revtex4}
\usepackage{bm,graphicx,subfigure,enumerate,latexsym}

\begin{document}
\title{Multiscale modeling and simulation for polymer melt flows between parallel plates}
\author{Shugo Yasuda
\footnote{Electronic mail: yasuda@cheme.kyoto-u.ac.jp}}
\author{Ryoichi Yamamoto
\footnote{Electronic mail: ryoichi@cheme.kyoto-u.ac.jp}}
\affiliation{
Department of Chemical Engineering, 
Kyoto University, Kyoto 615-8510, Japan
and
CREST, Japan Science and Technology Agency, Kawaguchi 332-0012, Japan.
}
\date{\today}

\begin{abstract}
The flow behaviors of polymer melt composed of short chains with ten beads between parallel plates are simulated by using a hybrid method of 
molecular dynamics and computational fluid dynamics.
Three problems are solved: creep motion under a constant shear stress and its recovery motion after removing the stress, pressure-driven flows, and the flows in rapidly oscillating plates.
In the creep/recovery problem, the delayed elastic deformation in the creep motion and evident elastic behavior in the recovery motion are demonstrated.
The velocity profiles of the melt in pressure-driven flows are quite different from those of Newtonian fluid due to shear thinning. 
Velocity gradients of the melt become steeper near the plates and flatter at the middle between the plates as the pressure gradient increases and the temperature decreases.
In the rapidly oscillating plates, the viscous boundary layer of the melt is much thinner than that of Newtonian fluid due to the shear thinning of the melt.
Three different rheological regimes, i.e., the viscous fluid, visco-elastic liquid, and visco-elastic solid regimes, form over the oscillating plate according to the local Deborah numbers.
The melt behaves as a viscous fluid in a region for $\omega\tau^R\lesssim 1$, and the crossover between the liquid-like and solid-like regime takes place around $\omega\tau^\alpha\simeq 1$ (where $\omega$ is the angular frequency of the plate and $\tau^R$ and $\tau^\alpha$ are Rouse and $\alpha$ relaxation time, respectively).
\end{abstract}


\maketitle

\section{Introduction}
Many products in our daily life contain soft matters (e.g., polymeric liquids, colloidal dispersion, and liquid crystals).
Soft matters involve complex internal degrees of freedom, such as orientation of molecules, dispersed particles, and phase-separated structure, and they exhibit peculiar flow behaviors coupled to the micro-scale dynamics, e.g., visco-elastic flow, shear thinning or thickening, and flow-induced phase transition.
Investigation of the complicated flow behaviors of soft matters are of great importance in various science and engineering fields, such as fluid mechanics, materials science, biological science, mechanical engineering, and chemical engineering.
To predict the flow behaviors of soft matters by computer simulation is challenging from both an academic and a practical point of view.
The difficulty in simulating soft-matter flows also is also due to the complicated couplings between the microscopic dynamics of internal degrees of freedom and macroscopic flow behaviors.
In the present paper, we demonstrate that multiscale modeling is a promising candidate for soft-matter simulations.
We solve the complicated visco-elastic flow behaviors of polymer melt by using a multiscale hybrid simulation method.

Usually, when computer simulations of polymer melt flows are performed, either computational fluid dynamics (CFD) or molecular dynamics (MD) are employed.
In the case of CFD, mechanical properties of fluids must be modeled mathematically in advance as a form of ``constitutive relationship'' to be used in simulations. 
CFD is thus valid only for the cases in which mechanical properties of fluids are not too complex.
Polymer melts, however, have very complicated mechanical properties in general.
In the case of MD simulation, in contrast, fluids consist of huge numbers of molecules of arbitrary shapes.
In principle, MD simulation is thus applicable for any flows of any complex fluids.
However, the drawback here is that enormous computational time is required to resolve the dynamics of all the constituent molecules.
Hence, MD simulation is not yet applicable to problems which concern large-scale motions far beyond molecular size, as is done in the present paper.
In order to overcome the weaknesses of the individual methods, we have developed a multi-scale simulation method that combines MD and CFD.

In our hybrid simulation method, the macroscopic flow behaviors are solved using a CFD scheme, but, instead of using any constitutive equations, a local stress is calculated by using a non-equilibrium MD simulation associated with each lattice node of the CFD simulation.
The basic idea of this type of hybrid simulation method was first proposed by E and Engquist\cite{art:03EE, art:07EELRV}, where the heterogeneous multiscale method (HMM) is presented as a general methodology for the efficient numerical computation of problems with multiscale problems.
The HMM has also been applied to the simulations of complex fluids.\cite{art:05RE}
The equation-free multiscale computation proposed by Kevrekidis {\it et al}. is also based on a similar idea and has been applied to various problems.\cite{art:03KGHKRT, art:09KS}
De {\it et al}. have developed a hybrid method, which is called scale bridging method in their paper, that can correctly reproduce the memory effect of a polymeric liquid and demonstrate the non-linear visco-elastic behaviors of polymeric liquid between oscillating plates.\cite{art:06DFSKK}
The methodology of the present hybrid simulation is the same as that of the scale bridging method.
The present hybrid method is also similar to the CONNFFESIT approach\cite{art:93LO, art:95FLO, art:97LPO} in that the macroscopic local stresses are obtained from the configurations of polymer molecules.
In the present method, however, we couple a CFD simulation with an MD simulation which involves full interactions of each molecule rather than the stochastic dynamics of single model polymers lacking any direct interactions between molecules.
Thus the present method can reproduce flow behaviors caused by the many-body effect, which is quite important, especially in the glassy polymers.
In previous work\cite{art:08YY, art:09YY}, we investigated the efficiency and accuracy of the hybrid method. 
We also clarified the property of noise arising in the hybrid method by comparing the results of the hybrid method with those from fluctuating hydrodynamics.
In addition, we analyzed the complicated rheological properties of a supercooled polymer melt in the viscous diffusion layer arising over the rapidly oscillating plate by using the scale bridging method.

In the present paper, various flow behaviors of polymer melts between the parallel plates are simulated using the scale bridging method.
Creep motion under a constant shear stress and its recovery motion in the stress-free state, pressure-driven flows, and flows in rapidly oscillating plates are investigated.
In the creep/recovery simulation, the evident elastic motion of polymer melt is demonstrated.
The result is also compared with that from a model constitutive relation.
For pressure-driven flow, the velocity profiles of melts with various temperatures and pressure gradients are demonstrated.
The shear-thinning behaviors of melts are also investigated.
In the problem of oscillating plates, the viscous boundary layer arising over the oscillating plate and rheological properties of the boundary layer are investigated for various temperatures, frequencies of the plate, and amplitudes of strain of the system.
The parameter sets used in the present paper are different from those used in the previous paper\cite{art:09YY}; in the previous paper, the amplitude of velocity of the plate is specified as a parameter instead of the amplitude of strain of the system in the present paper.

For the problems of pressure-driven flow and oscillating plates, the velocity profiles are quite non-uniform between the plates, and two different characteristic length scales appear in each problem that must be resolved: one is that of a polymer chain, which is the scale of the MD simulation, and the other is that of the flow behaviors of the melt, e.g., the width between the plates or the thickness of boundary layer arising over the oscillating plate, which is the scale of the CFD simulation.
These problems constitute important applications of multiscale modeling; it is quite difficult to solve each problem by using a full MD simulation because the width of the plates and thickness of the boundary layer are much larger than the size of a polymer chain.

In the following text, the multiscale modeling and simulation method for the present problem are stated in Sec. II.
Some model constitutive relations for polymeric liquids and the difficulties with using these in CFD simulations are also mentioned in Sec. II.
In Sec. III, the simulation results for the creep and recovery, pressure driven flows, and oscillating plates are given.
The rheological properties are also discussed in Sec. III.
Finally, a summary and an outlook for the hybrid simulations are given in Sec. IV.

\section{Multiscale modeling and simulation method}
\begin{figure*}[t]
\includegraphics[scale=1]{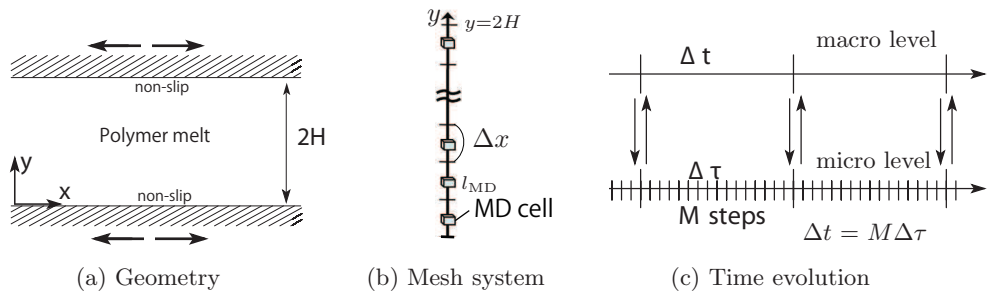}
\caption{
Schematics for geometry of problem, mesh system, and time-evolution scheme.
}\label{fig1}
\end{figure*}

We consider the polymer melt with a uniform density $\rho_0$ and a uniform temperature $T_0$ between two parallel plates (see Fig. \ref{fig1}(a)). 
The upper- and lower- plate can move in the $x$-direction.
The melt is composed of short chains with ten beads.
The number of bead particles composing each chain in the MD simulation is represented by $N_{\rm b}$.
Thus $N_{\rm b}=10$.
All of the bead particles interact with a truncated Lennard-Jones potential defined by\cite{art:90KG},
\begin{equation}\label{eq1}
U_{\rm LJ}(r)=
\left\{
\begin{array}{c c}
4\epsilon\left[
({\sigma}/{r})^{12}
-({\sigma}/{r})^{6}
\right]
+\epsilon & (r\le 2^{1/6}\sigma),\\
0 & ( r> 2^{1/6}\sigma).
\end{array}
\right .
\end{equation}
By using the repulsive part of the Lennard-Jones potential only, we may prevent spatial overlap of the particles.
Consecutive beads on each chain are connected by an anharmonic spring potential,
\begin{equation}\label{eq2}
U_{\rm F}(r)=-\frac{1}{2}k_c R_0^2 \ln
\left[
1-({r}/{R_0})^2
\right],
\end{equation}
where $k_c$=30$\epsilon/\sigma^2$ and $R_0$=$1.5\sigma$.
The number density of the bead particles is fixed at $\rho_0/m$=1/$\sigma^3$, where $m$ is the mass of the bead particle.
With this number density the configuration of bead particles becomes severely jammed at a low temperature, resulting in the complicated non-Newtonian viscosity and long-time relaxation phenomena typically seen in glassy polymers.\cite{book:92M, art:02YO}

We assume that macroscopic quantities are uniform in the $x$- and $z$-directions, $\partial/\partial x$=$\partial/\partial z$=0.
The macroscopic velocity $v_\alpha$ is described by the following equations,
\begin{equation}\label{eq3}
\rho_0\frac{\partial v_x}{\partial t} = 
\frac{\partial \sigma_{xy}}{\partial y},
\end{equation}
and $v_y$=$v_z$=0, where $t$ is the time and $\sigma_{\alpha\beta}$ is the stress tensor. 
For pressure-driven flow, $-\Delta P$ is also added to the right-hand side of Eq. (\ref{eq3}), where $\Delta P$ is a uniform pressure gradient in $x$-direction.
Here and afterwards, the subscripts $\alpha$, $\beta$, and $\gamma$ represent the index in Cartesian coordinates, i.e., \{$\alpha$,$\beta$,$\gamma$\}=\{$x$,$y$,$z$\}.
We also assume the non-slip boundary condition on each plate.

The constitutive relation of the stress tensor may be written in a functional of the history of velocity gradients on the fluid element, 
\begin{equation}\label{eq4}
\sigma_{\alpha\beta}(t,x_\alpha)=F_{\alpha\beta}[\kappa_{\alpha\beta}(t',x_\alpha'(t'))],
\quad {\rm with} \quad t'\le t,
\end{equation}
where $\kappa_{\alpha\beta}$ is the velocity gradient, $\kappa_{\alpha\beta}=\partial v_\alpha/\partial x_\beta$, and $x'_\alpha(t')$ represents the path line along which a fluid element has been moving.
Note that the temporal value of the stress tensor of a fluid element depends upon the previous values of the velocity gradients of the fluid element.
In the one-dimensional problem, however, the equation simplifies to a functional involving only the local strain rate $\dot \gamma$=$\partial v_x/\partial y$ since the macroscopic velocity is restricted to the $x$-direction where the macroscopic quantities are assumed to be uniform;
\begin{equation}\label{eq5}
\sigma_{xy}(t,y)=F_{xy}[\dot \gamma(t',y)],
\quad{\rm with}\quad t'\le t.
\end{equation}
Constructing a model for the constitutive relation is one of the most important goals in complex fluid research, and many models have been proposed.\cite{book:87BAH,book:88L}
The simplest model for the visco-elastic motion is one proposed by Maxwell\cite{art:1867M}, which is written as
\begin{equation}\label{eq_6}
\sigma_{xy}+\lambda_1\frac{\partial \sigma_{xy}}{\partial t}
=\eta_1\dot\gamma.
\end{equation}
where $\lambda_1$ is a time constant and $\eta_1$ a viscosity.
For the steady state, this equation simplifies to the Newtonian fluid with viscosity $\eta_1$, while for sudden changes in stress, the time derivative term dominates the left-hand side of the equation, and then integration with respect to time gives the Hookean solid with elastic modulus $G=\eta_1/\lambda_1$.
The model which includes the time derivative of $\dot \gamma$ in the right-hand side of Eq. (\ref{eq_6}) has also been proposed by Jeffreys,\cite{book:76J}
\begin{equation}\label{eq_7}
\sigma_{xy}+\lambda_1\frac{\partial \sigma_{xy}}{\partial t}
=\eta_1\left(\dot\gamma+\lambda_2\frac{\partial \dot \gamma}{\partial t}\right).
\end{equation}
This equation contains two time constants $\lambda_1$ (the ``relaxation time'') and $\lambda_2$ (the ``retardation time'').
This model may reproduce the delayed elastic motion for the sudden change of stress.
Models also exist for the multiple relaxation modes and nonlinear simulations involving either the shear-rate dependent viscosity $\eta_1(\dot \gamma)$ or the nonlinear stress term.\cite{art:63WM,art:82G}
Usually, when one performs the CFD simulations for the polymer melt flows, one chooses an appropriate model of them empirically according to the physical properties of the problem or the computational convenience.
The time constants, viscosities, and their dependence on the shear rate in the equation must be specified in advance.
However, no systematic methods exist to choose a suitable model for each problem, and determination of time constants, viscosities, and their dependence on the shear rate is quite difficult, especially for glassy materials.
In our multiscale modeling, instead of using explicit formulas for the constitutive relation, the local stress is generated by the non-equilibrium MD simulation associated with each local point. 
See Fig. \ref{fig1} (b).

We use a common finite volume method with a staggered arrangement for the CFD calculation, where the velocity is computed at the node of each slit and the stress is computed at the center of each slit.
For the time-integration scheme, we use the simple explicit Euler method with a small time-step size $\Delta t$.
The local stress is calculated  at each time step of CFD by using the non-equilibrium MD simulation with a small cubic MD cell associated with each slit according to a local strain rate.
In each MD simulation, we solve the so-called SLLOD equations of motion with the Gaussian iso-kinetic thermostat:\cite{book:89AT, book:08EM}
\begin{subequations}\label{eq_8}
\begin{equation}\label{eq_8a}
\frac{d {\bm R}^k_j}{dt}=
\frac{{\bm p}^k_j}{m}+\dot \gamma {R_y}^k_j {\bm e}_x,
\end{equation}
\begin{equation}\label{eq_8b}
\frac{d {\bm p}^k_j}{dt}=
{\bm f}^k_j - \dot \gamma {p_y}^k_j {\bm e}_x
-\zeta {\bm p}^k_j,
\end{equation}
\end{subequations}
where ${\bm e}_x$ is the unit vector in the $x$ direction and the indexes $k$ and $j$ represent the $k$th polymer chain ($k=1,\cdots,N_{\rm p}$) and the $j$th bead  ($j=1,\cdots,N_{\rm b}$) on each chain, respectively.
Here the number of polymer chains contained in each MD cell is represented by $N_{\rm p}$.
${\bm R}^k_j$ and ${\bm p}^k_j + m\dot\gamma {R_y}^k_j{\bm e}_x$ are the position and momentum of the $j$th bead on the $k$th polymer chain, respectively, 
${\bm f}^k_j$ is the force acting on the $j$th bead on the $k$th polymer due to the potentials described in Eqs. (\ref{eq1}) and (\ref{eq2}), and $\dot\gamma$ is the shear rate subjected on each MD cell, which corresponds to the local shear rate at each slit of the CFD calculation.
Note that, in the SLLOD equations, ${\bm p}^k_j/m$ represents the deviation of velocity of each particle from the mean flow velocity $\dot\gamma {R_y}^k_j{\bm e}_x$ in the MD cell.
The friction coefficient $\zeta$ is determined to satisfy the constant temperature condition $dT/dt=0$ with $T=\sum_{j,k} |{\bm p}^k_j|^2/3mN_{\rm p}N_{\rm b}$, where $\sum_{j,k}$ represents the summation all over the bead particles in each MD cell.
The friction coefficient $\zeta$ is calculated as
\begin{equation}\label{eq_9}
\zeta=\sum_{j,k}({\bm f}_j^k\cdot{\bm p}^k_j-\dot\gamma {p_x}^k_j {p_y}^k_j)/\sum_{j,k}|{\bm p}^k_j|^2,
\end{equation}
We integrate Eq. (\ref{eq_8}) with Eq. (\ref{eq_9}) by using the leapfrog algorithm.\cite{art:84BC}
The space integral of the microscopic stress tensor reads as
\begin{align}\label{eq_10}
\Pi_{\alpha\beta}(t)=&
\frac{1}{m}\sum_{k=1}^{N_{\rm p}}\sum_{j=1}^{N_{\rm b}}{p_\alpha}^k_j{p_\beta}^k_j
-\sum_{\rm all pairs} \frac{dU_{\rm LJ}(\xi)}{d\xi}\frac{\xi_\alpha\xi_\beta}{\xi}
\nonumber\\
&-\sum_{k=1}^{N_{\rm p}}\sum_{j=1}^{N_{\rm b}}\frac{dU_{\rm F}(\xi)}{d\xi}\frac{\xi_\alpha\xi_\beta}{\xi},
\end{align}
where we rewrite the momentum of the $j$th bead on the $k$th chain, ${\bm p}^k_j + m\dot\gamma {R_y}^k_j{\bm e}_x$, as ${\bm p}_j^k$.
$\bm \xi$ in the right-hand side of Eq. (\ref) represents the relative vector ${\bm R}^k_j - {\bm R}^{k'}_{j'}$ between the two beads, ${\bm R}^k_j$ and ${\bm R}^{k'}_{j'}$, in the second term
and the relative vector ${\bm R}^k_j - {\bm R}^{k}_{j+1}$ between the two consecutive beads on the same chain, ${\bm R}^k_j$ and ${\bm R}^{k}_{j+1}$, in the third term.

In the present problem we cannot assume a local equilibrium state at each time step of the CFD simulation since the relaxation time of the stress may become much longer than the time-step size of the CFD simulation (in which the macroscopic motions of the system should be resolved).
In the current simulations, the simple time-averages of the temporal stresses of the MD (averaged over the duration of a time-step of the CFD simulation) are used as the stresses at each time step of the CFD calculation without ignoring the transient time necessary for the MD system to be in steady state. See Fig. \ref{fig1}(c).
Thus the time integration of the macroscopic local stresses $\bar\sigma_{\alpha\beta}$ are calculated with the microscopic stress tensor Eq. (\ref{eq_10}) as
\begin{align}\label{eq_11}
\bar\sigma_{\alpha\beta}(t,y)&=\int_t^{t+\Delta t}\sigma_{\alpha\beta}(t',y)dt'
\nonumber\\
&=\frac{1}{l_{\rm MD}^3 }\int_{t}^{t+\Delta t}\Pi_{\alpha\beta}(\tau;\dot\gamma(t,y))d\tau,
\end{align}
where $\Delta t$ is the time-step size of the CFD simulation and $l_{\rm MD}$ is the side length of the cubic MD cell.
Note that the second argument of $\Pi_{\alpha\beta}$ in Eq. (\ref{eq_11}), i.e., $\dot \gamma(t,y)$, is constant in the integral interval.
This indicates that the shear rate to which each MD cell is subjected is also constant over a duration $\Delta t$ in each MD simulation.
The final configuration of molecules obtained at each MD cell is retained as the initial configuration for the MD cell at the next time step of the CFD.
Thus we trace all temporal evolutions of the microscopic configurations with a microscopic time step so that the memory effects can be reproduced correctly. 
Note that compared to a full MD simulation, we can save computation time with regard to the spatial integration by using MD cells that are smaller than the slit size used in the CFD simulation. 
The efficiency of the performance of the present hybrid simulation is represented by a saving factor defined by the ratio of the slit size used in the CFD simulation $\Delta x$ to the cell size of the MD simulation $l_{\rm MD}$, $\Delta x$/$l_{\rm MD}$.
It also should be noted that, in addition to the saving factor $\Delta x/l_{\rm MD}$, the present hybrid method is quite suitable as a parallel computational algorithm since the MD simulations associated with each mesh of the CFD, which cost a large part of the total simulation time, are performed independently.

We solve the polymer melt flows in the following three problems by using the hybrid method; (i) creep motion under a constant shear stress and recovery after removing the stress, (ii) pressure-driven flows and (iii) flows in rapidly oscillating plates.
In the first problem, the simple visco-elastic behavior of the melt is demonstrated in order to verify that the present hybrid method can reproduce the visco-elastic motion correctly.
In the second problem, we demonstrate the peculiar velocity profile of the melt produced due to shear thinning near the plate.
We also investigate the local rheology and the microscopic configurations at local points.
In the third problem, the viscous boundary layers arising near the oscillating plates are resolved.
The macroscopic quantities are quite non-uniform in the boundary layers, and the local rheological properties are also changed drastically according to the local flow fields.

\section{Simulation results}
Hereafter, the quantities normalized by the units of length $\sigma$ and time $\tau_0$=$\sqrt{m\sigma^2/\epsilon}$ are denoted with a hat ``$\Hat{\quad}$''. 
In the following simulations, we fix the time-step size of the CFD simulation $\Delta t$, sampling duration of the MD simulation $t_{\rm MD}$ and time-step size of the MD simulation $\Delta \tau$ as $\Delta \hat t$=$\hat t_{\rm MD}$=1 and $\Delta \hat \tau$=0.001, respectively. 
Thus, 1000 MD steps ($M=1000$ in Fig. \ref{fig1}(c).) are performed in each MD cell at each time step of the CFD computation. 
One hundred chains with ten beads are confined in each cubic MD cell with a side length $\hat l_{\rm MD}$=10; thus $N_{\rm p}$=100 and $N_{\rm b}$=10.
The density of the melt is fixed to $\hat \rho$=1.
The temperature of the melt is $\hat T_0$=0.2 in the creep and recovery problem, and $\hat T_0$=0.2 and 0.4 are considered in the problems of pressure-driven flow and oscillating plates.
At this number density and a low temperature, the polymer melt involves complicated non-Newtonian rheology.

\begin{figure*}[t]
\includegraphics[scale=1]{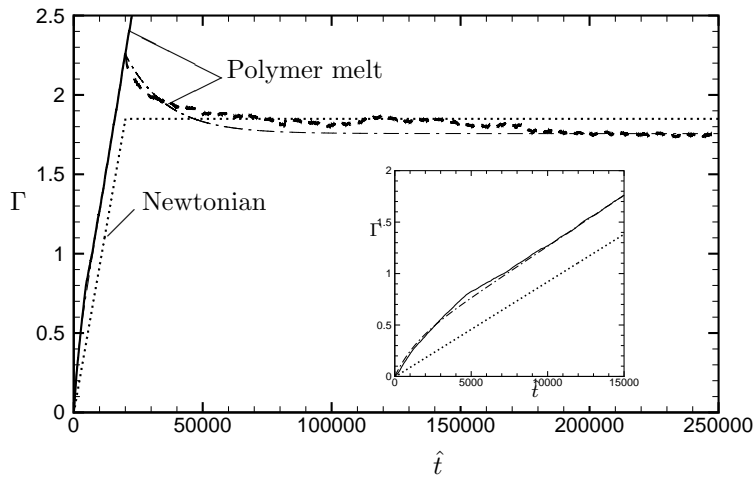}
\caption{
The time evolution of the strain of the system, $\Gamma=\left [ u_x(y=2H)-u_x(y=0) \right ]/2H$.
The solid line shows the creep motion of the melt.
The dashed line shows the recovery motion of the melt after removing the stresses on the plates at $\hat t$=20000.
The dotted line shows the result for the Newtonian fluid.
The dash-dotted line shows the result given by the Jeffreys model in Eq. (\ref{eq_7}).
The inset shows the time evolution of the strain in the beginning of creep motion.
}\label{fig2}
\end{figure*}
\subsection{Creep and Recovery}
\begin{figure}[t]
\includegraphics[scale=1]{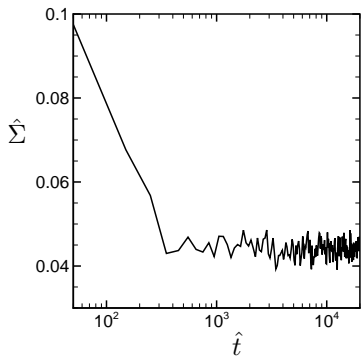}
\caption{
The time evolution of the standard deviation between the local stresses and the subjected stress on the plates $\sigma_0$ described in Eq. (\ref{eq_16}).
}\label{fig3}
\end{figure}

The upper and lower plates were each subjected to a constant shear stress $\sigma_0$ at a time $t$=0.
For a positive value of $\sigma_0$, the upper plate slides right and the lower plate left in Fig. \ref{fig1}(a).
We set the shear stress $\sigma_0$ and the distance between the plates $2H$ as $\hat \sigma_0$=0.1 and $2\hat H$=800, respectively.
The distance between the plates $2\hat H$ is divided into seventeen slits.
The present mesh system is a little different from that in Fig. \ref{fig1}(b).
In the present mesh system, sixteen MD cells are used to sample the local stresses; these cells are associated with the nodes of the slits and the local velocities are calculated at the centers of each slit.
Thus the saving factor in the present simulation is $\Delta x/l_{\rm MD}=4.7$.
The stress to which the plate is subjected is transmitted inside the material immediately, and after a short transient time, the local macroscopic quantities of the melt are in uniform states between the plates (although they fluctuate due to the noise arising from each MD cell).
The transient time $\delta t$ may be estimated from Fig. \ref{fig3} as $\delta \hat t \sim $ 500.
The system is deformed uniformly after a short transient time.

We measure the strain of the system $\Gamma(t)$ as 
\begin{equation}\label{eq_12}
\Gamma(t)=\frac{u_x(y=2H,t)-u_x(y=0,t)}{2H},
\end{equation}
where $u_x(y,t)$ is the displacement of the melt in $x$-direction, which is calculated as
\begin{equation}\label{eq_13}
u_x(y,t)=\int_0^t v_x(y,\tau)d\tau.
\end{equation}
Figure \ref{fig2} shows the time evolution of the strain of the system $\Gamma(t)$.
In the figure, the solid line shows the creep motion and the dashed line shows the recovery motion after removing the stresses on the plates at a time $\hat t$=20000.
The inset zooms in on the creep motion.
It can be seen that, in the creep motion, the strain $\Gamma$ increases rapidly at the beginning, say $\hat t\simeq[0,5000]$, and, as time passes, the time evolution of the strain becomes linear.
The rapid evolution of the strain at the beginning of the creep motion is caused by the elasticity of the meld.
However, since it is different from the pure elastic deformation, in which a finite strain is instantaneously obtained when subjecting the stress, the strain of the melt evolves rather continuously.
This behavior demonstrates the delayed elastic deformation of the melt.
The linear evolution of the strain corresponds to the viscous flow.
The viscosity of the melt in the linear evolution regime $\eta_1$ is calculated from the slope, which reads $\hat \eta_1$=1007.
The result of the Newtonian fluid with the viscosity $\hat \eta_1$ is also shown in Fig \ref{fig2} by the dotted line for comparison.
When removing the stress on the plate (the dashed line in Fig. \ref{fig2}),  the strain of the system starts to decrease very slowly to the line of the Newtonian fluid.
This recovery motion of strain demonstrates the evident elastic motion; the elastic strain of the melt stored in the creep motion is being released after removing the stresses on the plates.

The delayed elastic deformation may be described using the Jeffreys model shown in Eq. (\ref{eq_7}).
By integrating Eq. (\ref{eq_7}) with $\sigma_{xy}=\sigma_0\Theta(t)$, where $\Theta(t)$ is the step function, the time evolution of the strain in the creep motion is written as follows:
\begin{equation}\label{eq_14}
\Gamma(t)=\frac{\sigma_0}{G_1}\left(1-e^{-\frac{t}{\lambda_2}}\right)
+\frac{\sigma_0}{\eta_1}t,
\end{equation}
where $G_1=\eta_1/(\lambda_1-\lambda_2)$.
The first term in the right-hand side of Eq. (\ref{eq_14}) represents the delayed elastic deformation and the second term the viscous flow.
By fitting our result to Eq. (\ref{eq_14}), the viscosity $\eta_1$, elastic modulus $G_1$, and retardation time $\lambda_2$ are estimated as $\hat \eta_1$=1007, $\hat G_1$=0.38, and $\hat \lambda_2$=13003.
The result of the Jeffreys model with these parameter values is shown in Fig. \ref{fig2} (inset) as a dash-dotted line.
It is evident that the creep motion obtained by the hybrid simulation can be fitted well by the delayed elastic motion described by the Jeffreys model Eq. (\ref{eq_14}).
The recovery motion for the Jeffreys model is written as 
\begin{equation}\label{eq_15}
\Gamma(t)=\sigma_0/G_1\left(1-e^{-\frac{t'}{\lambda_2}}\right)e^{-\frac{t-t'}{\lambda_2}}+\Gamma_\infty,
\end{equation}
where $\Gamma_\infty$ is a elastic strain remaining at an infinite time.
Eq. (\ref{eq_15}) is obtained by integrating the strain rate $\dot \Gamma(t)$ (for $t>t'$), which is obtained by integrating Eq. (\ref{eq_7}) with $\sigma_{xy}=\sigma_0[1-\Theta(t-t')]$ using $\dot \Gamma(t_0)$ for Eq. (\ref{eq_14}), from a time $t$ to the infinity $t=\infty$.
In the recovery motion, however, the parameter values in the Jeffreys model differ from those in the creep motion since the mechanical property of the melt changes according to the state of motion.
The result by the hybrid simulation in the recovery motion may be fitted to Eq. (\ref{eq_15}) with $\hat G_1=0.14$, $\hat \lambda_2=1.6\times 10^4$, $t'=20000$, and $\Gamma_\infty$=1.76 which is a value of strain obtained by the hybrid simulation at $t=2.5\times 10^4$.
The result of the Jeffreys model with these parameter values in recovery motion is shown as the dash-dotted line in Fig. \ref{fig2}.

As previously mentioned, the local macroscopic quantities fluctuate due to the noises arising from each MD cell.
These fluctuations can be measured by the standard deviation of the local stresses from the subjected stress on the plates, which is defined by
\begin{equation}\label{eq_16}
\Sigma(t)=\sqrt{\int_{-b/2}^{b/2} \int_0^H (\sigma_{xy}(y', t +\tau') - \sigma_0)^2 dy' d\tau'\Big / Hb}.
\end{equation}
The time evolution of the standard deviation $\Sigma$ with $\hat b$=100 is shown in Fig. \ref{fig3}.
After a short transient time $\delta t$, $\delta \hat t\sim$ 500, the local stresses fluctuate around the uniform state with a subjected stress on the plates $\sigma_0$.
The standard deviation in the fluctuating states is estimated to be approximately $\hat \Sigma \simeq 0.045$.

\subsection{Pressure-driven flow}
We consider a uniform pressure gradient in the $x$-direction $\Delta P$.
Both the upper- and lower-plates are at rest.
For the present problem, Eq. (\ref{eq3}) is modified to include the uniform pressure gradient; $-\Delta P$ is added in the right-hand side of Eq. (\ref{eq3}).
The distance between the plates is set as $2\hat H=1600$.
The mesh system is the same as that shown in Fig. {\ref{fig1}}(b).
We use thirty-two slits in the distance $2H$.
The saving factor in this case is $\Delta x/l_{\rm MD}=5$.
Hybrid simulations are performed for the melts using various parameter values and are shown in Table \ref{table1}.
\begin{table}[t]
\caption{
Parameter values for pressure-driven flows and simulation times.
$T$ is the temperature and $\Delta P$ is the pressure gradient of the melt.
$t_{\rm Max}$ is the simulation time for each case.
}\label{table1}
\begin{tabular}{c cc c c c }
\hline\hline
      && $\hat T_0$ & $\Delta\hat  P\times 10^4$ & $\hat t_{\rm Max}/10^4$\\
\hline
Case A&& 0.2      & 1  & 40 \\
Case B&& 0.2      & 3  & 20 \\
Case C&& 0.2      & 5  & 10 \\
Case D&& 0.4      & 3  &  7 \\
Case E&& 0.4      & 5  &  7 \\
\hline\hline
\end{tabular}
\end{table}

\begin{figure*}[t]
\includegraphics[scale=1]{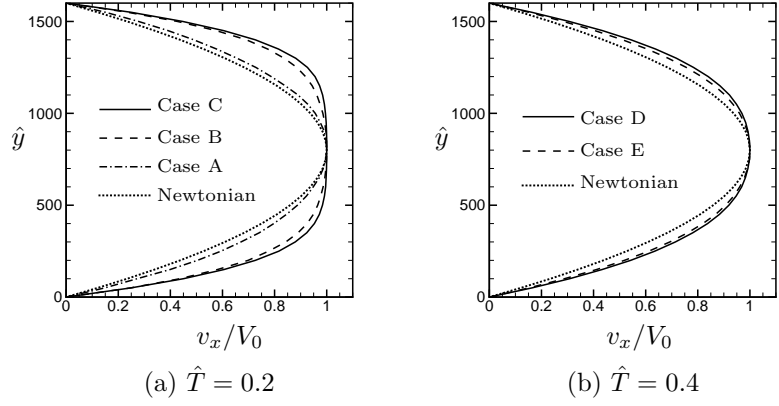}
\caption{
Comparison of the normalized velocity profiles in the steady states of the melt to those of the Newtonian fluid for the temperature $\hat T=0.4$ [for (a)] and $\hat T=0.2$ [for (b)].
$V_0$ is the velocity at the middle between the plates.
The value of $V_0$ for each flow is shown in Table \ref{table2}.
Note that the normalized velocity profile of the Newtonian fluid does not depend on either the pressure gradient $\Delta P$ or the temperature $T$.
}\label{fig4}
\end{figure*}
\begin{table}[b]
\caption{
The velocity of the melt at the middle between the plates $V_0$ in pressure-driven flows.
For the Newtonian fluid with a viscosity $\nu_0$, $V_0$ is written as $V_0=-(\Delta P\rho_0/2\nu_0)H^2$.
}\label{table2}
\begin{tabular}{cc c cc}
\hline\hline
\multicolumn{2}{c}{$\hat T=0.2$}&&\multicolumn{2}{c}{$\hat T=0.4$}\\
\cline{1-2}\cline{4-5}
$\Delta \hat P\times 10^4$ & $\hat V_0$ &&$\Delta \hat P\times 10^4$ & $\hat V_0$\\
\hline
1 & 0.024 && 3 & 1.67\\
3 & 0.31  && 5 & 3.90\\
5 & 1.84  && \multicolumn{2}{c}{---}\\
\hline\hline
\end{tabular}
\end{table}
Figure \ref{fig4} shows the comparison of the velocity profiles normalized by the velocities at the middle between the plates $V_0$, $V_0=v_x(y=H)$, for the melt and for the Newtonian fluid.
The velocity profiles of the melt with $\hat T$=0.4 are time-averaged over $\hat t=[50001,70000]$
and those of the melt with $\hat T$=0.2 are time-averaged over
$\hat t=[200001,400000]$ for $\Delta \hat P$=$1\times 10^{-4}$,
$\hat t=[100001,200000]$ for $\Delta \hat P$=$3\times 10^{-4}$,
and $\hat t=[50001,100000]$ for $\Delta \hat P$=$5\times 10^{-4}$.
The value of $V_0$ for each flow is also shown in Table \ref{table2}.
It can be seen that the velocity profiles of the melts are quite different from those of the Newtonian fluid.
As the pressure gradient increases, the velocity gradients of the melts become steeper near the plates and are more gradual (or rather a plateau for $\hat T=0.2$) in the middle region.
This feature is enhanced at a low temperature $\hat T=0.2$.
The dependence of the velocity at the middle between the plates $V_0$ on the pressure gradient $\Delta P$ for the melt is quite different from that for the Newtonian fluid.
For the Newtonian fluid, the maximum velocity $V_0$ depends linearly on the pressure gradient, while for the polymer melt its dependence is quite non-linear.
Especially, at a low temperature $\hat T=0.2$, $V_0$ increases more than tenfold while the pressure gradient increases only threefold; $\hat V_0=0.024$ to 0.31 for $\Delta \hat P=1\times 10^{-4}$ to $3\times 10^{-4}$.

\begin{figure}[t]
\includegraphics[scale=1]{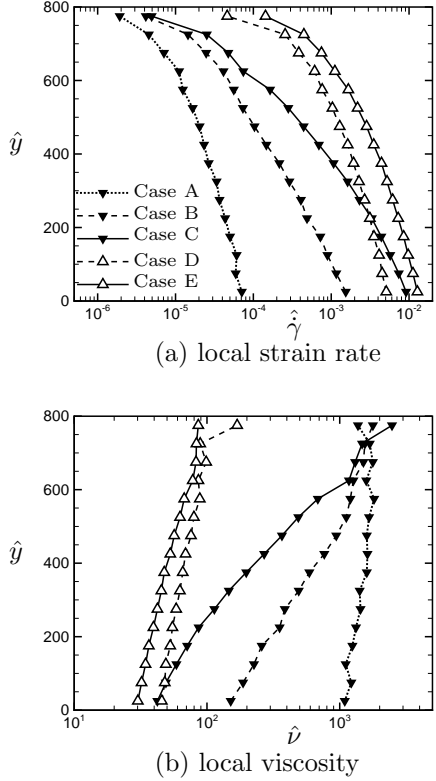}
\caption{
Spatial variation of the local strain rates [for (a)] and the local melt viscosities [for (b)].
The profiles in the lower-half space are plotted. The profiles in the upper-half space are antisymmetric for the strain rate and symmetric for the viscosity.
}\label{fig5}
\end{figure}
In Fig. \ref{fig5}, we show the spatial variations of the local strain rates and local viscosities of the melts.
Here, the local strain rates are time-averaged as in the velocity profiles.
The local viscosities are calculated by dividing the time averages of the local stresses by those of the local strain rates.
It is seen that the shear thinning occurs near the plate; i.e., the local viscosities becomes thinner as the local strain rates near the plate increase.
The shear thinning is enhanced at a low temperature and a high pressure gradient.
For a melt with a low temperature $\hat T$=0.2 and a high pressure gradient $\Delta \hat P$=5$\times 10^{-4}$, local viscosity  near the plate decreases to less than 3$\%$ of that in the middle.
Thus, the velocity gradient becomes quite large compared to that for the middle between the plates, so that the flatter velocity profile shown in Fig. \ref{fig4} is produced.

\begin{figure}[t]
\includegraphics[scale=1]{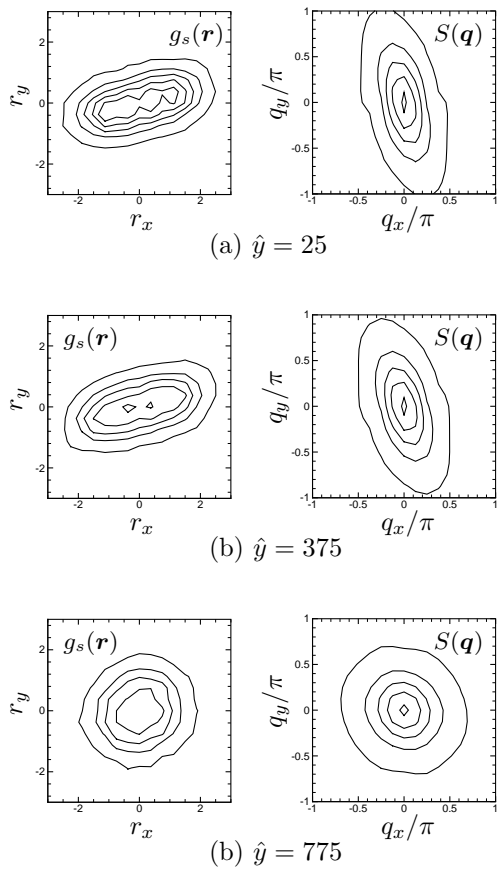}
\caption{
The local bead distribution function $g_s({\bm r})$ defined by Eq. (\ref{eq_17}) in the $r_x-r_y$ plane ($r_z=0$) and local structure factor $S({\bm{q}})$ defined by Eq. (\ref{eq_18}) in the $q_x-q_y$ plane ($q_z=0$) for the melt for Case B at $\hat y$=25 [for (a)], $\hat y$=375 [for (b)] and $\hat y$=775 [for (c)].
The contour lines show the values 0.01 + 0.02$k$ for $g_s$ and 0.05 + 0.2$k$ for $S$ with $k=0,1,\cdots,4$ from outer to inner.
}\label{fig6}
\end{figure}
The microscopic configurations of polymer chains are also investigated using the bead distribution function $g_s$ and structure factor $S$ defined below, which are calculated from the microscopic configurations of bead particles obtained at each MD cell.
The bead distribution function $g_s({\bm r})$ and structure factor $S({\bm q})$ are defined by
\begin{equation}\label{eq_17}
g_s({\bm r})=\frac{1}{{N_{\rm b}}}\sum_{j=1}^{N_{\rm b}}
\langle
\delta(\bm{R}^k_j-{\bm R}^k_G-{\bm r})
\rangle_k,
\end{equation}
with ${\bm R}^k_G={N_{\rm b}}^{-1}\sum_{i=1}^{N_{\rm b}} {\bm R}^k_i$, and
\begin{equation}\label{eq_18}
S({\bm q})=\frac{1}{{N_{\rm b}}^2}\sum_{i,j=1}^{N_{\rm b}}
\langle
\exp[{\rm i}{\bm q}\cdot(\bm{R}^k_i-{\bm R}^k_j)]
\rangle_k,
\end{equation}
where $k$ is the index to represent the $k$th polymer chain and $\bm R^k_i$ represents the position of the $i$th bead particle on the $k$th polymer chain in each MD cell.
Here, the assembled average $\langle \cdots \rangle_k$ is calculated by the average over all the configurations of polymer chains obtained at $\hat t=100000 + 1000n\,(n=0,\cdots, 100)$ in each MD cell.
In Fig. \ref{fig6}, we plot the bead distribution function $g_s$and structure factor $S$ for the case with $\hat T=0.2$ and $\Delta \hat P=3\times 10^{-4}$ (Case B) at different positions. 
It is seen that the coherent structure is produced near the plate, while the incoherent structure is produced in the middle between the plates.
The polymer chains are elongated in the flow direction ($x$-axis) near the plate, while in the middle between the plates bead particles are distributed rather randomly around the center of mass of each polymer chain.

\subsection{Oscillating plates}

\begin{table}[t]
\caption{
The parameter values for the problem of oscillating plates
and the thickness of the boundary layer produced in each case (which is also shown in Figs. \ref{fig5} and \ref{fig6}).
}\label{table3}
\begin{tabular}{c cc c c c cc}
\hline\hline
      && $\hat T_0$ & $2\pi/\hat \omega_0$ & $\Gamma_0$ & $\hat H$ && $\hat l_\nu$ \\
\hline
Case I&& 0.2      & $1024$    & 0.5              & 1600 &\,& 750\\
Case II&& 0.2      & $256$     & 0.5              & 800 && 200\\
Case III&& 0.4      & $256$    & 0.1             & 800  && 250\\
Case IV&& 0.2      & $256$     & 0.1              & 800 && 550\\
\hline\hline
\end{tabular}
\end{table}
The lower and upper plates begin to oscillate with a constant angular frequency $\omega_0$ at a time $t=0$ as, respectively, 
\begin{equation}\label{eq_19}
v_w = \pm v_0 \cos(\omega_0 t),
\end{equation}
with $v_0=\Gamma_0 \omega_0 H$.
Here $\Gamma_0$ represents the amplitude of strain of the system.
Thus, the strain of the system $\Gamma(t)$ in Eq. (\ref{eq_12}) is written as
\begin{equation}\label{eq_20}
\Gamma(t)=\Gamma_0\sin(\omega_0t).
\end{equation}
We perform the hybrid simulations for the parameters listed in Table \ref{table3}.
The distance between the plates $2\hat H$ is divided into sixty-four slits.
The mesh system is the same as that shown in Fig \ref{fig1}(b).
Thus sixty-four MD cells are associated with the centers of each slit.
The saving factor $\Delta x/l_{\rm MD}$ is 5 for Case I in Table \ref{table3} and 2.5 for Case II--IV in Table \ref{table3}.

\begin{figure}[t]
\includegraphics[scale=1]{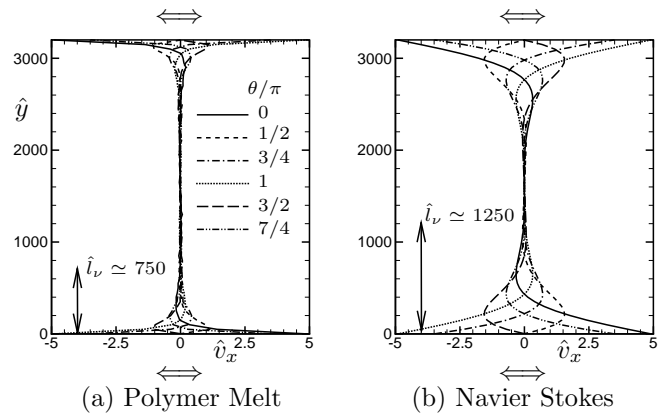}
\caption{
The snapshots of velocity profiles in oscillating plates at $\omega_0 t=50\pi + \theta$,
$\theta/\pi$=0, 1/2, 3/4, 1, 3/2, and 7/4, for $\hat \omega_0=2\pi/1024$ and $\Gamma_0$=0.5.
(a) The result for the polymer melt with $\hat T_0=0.2$ [Case I].
(b) The result for the Newtonian fluid.
The vertical axis shows the height $\hat y$ and the horizontal axis shows the velocity $v_x$.
$\hat l_\nu$ represents the thickness of the boundary layer, in which the amplitude of
the local oscillating velocity is more than 1 $\%$  of that of the oscillating plate, $|v_x|/v_0>0.01$.
}\label{fig7}
\end{figure}
\begin{figure*}[htbp]
\includegraphics[scale=1]{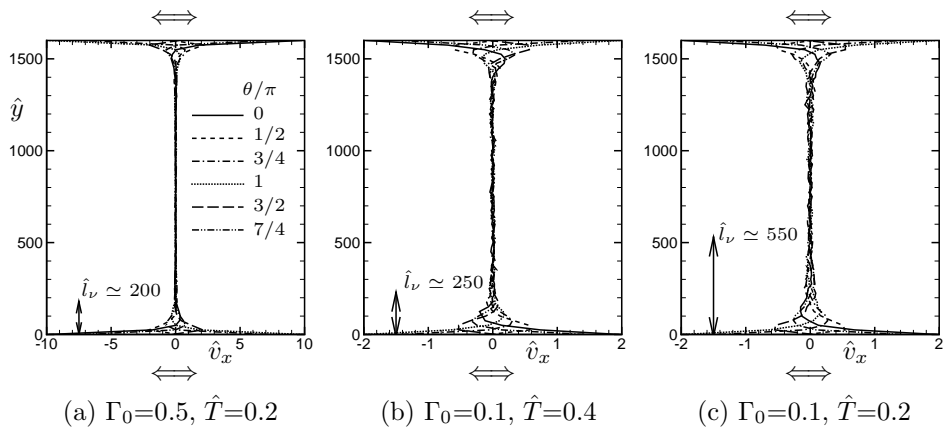}
\caption{
The snapshots of velocity profiles in oscillating plates at $\omega_0 t=100\pi + \theta$,
$\theta/\pi$=0, 1/2, 3/4, 1, 3/2, and 7/4, for $\hat \omega_0=2\pi/256$.
(a) The result for Case II, (b) that for Case III, and (c) that for Case IV.
See also the caption for Fig. \ref{fig5}.
}\label{fig8}
\end{figure*}
Figures \ref{fig7} and \ref{fig8} show the instantaneous velocity profiles between the oscillating plates.
The thickness of the viscous boundary layer near the oscillating plate $l_\nu$, which is defined by the specification that the amplitude of the local oscillating velocity is 1$\%$ of that of the oscillating plate, i.e., $|v_x(y=l_\nu)|/v_0=0.01$, is also shown in each figure.
Figure \ref{fig7} shows the comparison of the melt for Case I and the Newtonian fluid with a constant viscosity $\hat \nu=228$, which corresponds to the dynamics viscosity of the model polymer melt at $\omega_0=2\pi/1024$ for small strain rates.
The dynamics viscosity is calculated by $G''/\omega_0$ (where $G''$ is the loss modulus in the linear regime).
The velocity profiles are quite different from each other.
It is seen that the boundary layer of the melt is much thinner than that of the Newtonian fluid.
(The thickness of the boundary layer $l_\nu$ may be expressed as $l_\nu \propto \sqrt{\nu_0/\omega_0}$ for the Newtonian fluid.)
This is caused by shear thinning; the local loss modulus near the boundary is, as we see in Fig. \ref{fig10}, much smaller than that of the linear regime.
Figure \ref{fig8} shows the comparison of the velocity profiles for Cases II--IV in Table \ref{table3}.
The comparison of (a) and (c) shows the effect of the amplitude of the oscillating velocity and that of (b) and (c) shows the effect of the melt temperature.
The viscosity becomes smaller as the temperature or strain rate increase, and thus the boundary layers become thinner in Case II and Case III compared with Case IV.

We also measure the ``local'' visco-elastic properties in terms of the ``local'' storage modulus $G'$ and loss modulus $G''$.
It should be noted that the local macroscopic quantities oscillate with different phase retardations at each separate point.
The local moduli are calculated as follows:
The discrete Fourier transforms of the temporal evolutions of the strain $\gamma$, $\gamma(t,y)=\int_0^t\dot\gamma(t',y)dt'$, and shear stress $\sigma_{xy}$ during the steady oscillation states are performed, and are written as 
\begin{equation}\label{eq_21}
g_k^l = \frac{1}{N}\sum_{n=1}^N g_n^l e^{-{\rm i}2\pi (n-1)(k-1)/N}\quad(k=1,\dots,N),
\end{equation}
with $g^l_n$=$g(n\Delta t,l\Delta x)$ ($n$=1,...,$N$ and $l$=0,...,64) , where $g$ represents the strain or shear stress (e.g., $g$=$\gamma$ or $\sigma_{xy}$).
Hereafter the subscript $k$ represents the mode in Fourier space.
The time evolution of the local strain at $y$=$y^l$ may be expressed by using the Fourier coefficients for the mode of oscillation of the plate $k_0$, $k_0$=1+($\omega_0/2\pi$)$N$, as
\begin{equation}\label{eq_22}
\gamma^l(t) = |\gamma|^l\cos( \omega_0 t + \delta^l),
\end{equation}
with $|\gamma|^l$ = $\sqrt{ {\rm Re}(2\gamma_{k0}^l))^2+{\rm Im}(2{\gamma}_{k0}^l)^2}$ and $\delta^l$ = $\tan^{-1}({\rm Im}({\gamma}_{k0}^l)/{\rm Re}({\gamma}_{k0}^l))$.
The time evolution of the local shear stress for the mode $k_0$ can also be expressed as
\begin{equation}\label{eq_23}
\sigma_{xy}^l(t) = \sigma_1^l\cos(\omega_0 t + \delta^l) - \sigma_2^l\sin(\omega_0 t + \delta^l),
\end{equation}
where
\begin{subequations}\label{eq_24}
\begin{align}
\sigma_1^l & = {\rm Re}(2{\sigma}_{k_0}^l)\cos \delta^l + {\rm Im}(2 {\sigma}_{k_0}^l)\sin \delta^l,
\label{eq_24a}
\\
\sigma_2^l &= {\rm Im}(2 {\sigma}_{k_0}^l)\cos \delta^l - {\rm Re}(2 {\sigma}_{k_0}^l)\sin \delta^l.
\label{eq_24b}
\end{align}
\end{subequations}
Thus, the local storage modulus $G'$ and loss modulus $G''$ are obtained, respectively, as $G'(y^l)$=$\sigma_1^l/|\gamma|^l$ and $G''(y^l)$=$\sigma_2^l/|\gamma|^l$.
We note that, in the Fourier transformations, the signals for 3$\omega_0$ and 5$\omega_0$ are also detected, although their contributions are much smaller than that for $\omega_0$. 
In Table \ref{table4}, we show the amplitudes of the harmonic contributions at the frequency 3$\omega_0$ and 5$\omega_0$ for the local stresses.
$|\sigma^l_{xy}(\omega)|$ is the spector of local stress in the Fourier space, which is obtained by $|\sigma^l_{xy}(\omega)|=\sqrt{\sigma^l_1(k)^2+\sigma^l_2(k)^2}$ with $k=1+(\omega/2\pi)N$.
Here, $\sigma^l_{1,2}(k)$ are calculated via Eq. (\ref{eq_24}) by replacing $k_0$ with $k$.
Table \ref{table4} represents the amplitude of non-linear response with respect to the frequency.
It is seen that the non-linear response is enhanced in the boundary layer near the oscillating plate.

\begin{table}[htbp]
\caption{
The harmonic contributions of spectors of the local stress $|\sigma_{xy}^l(\omega)|$ for frequency $3\omega_0$ and $5\omega_0$ in Case I and II.
}\label{table4}
\begin{tabular}{cccc}
\hline\hline
\multicolumn{4}{c}{Case I}\\
\cline{1-4}
$\hat y$ 
& $|\sigma_{xy}(\omega_0)|$ 
& $\displaystyle\frac{|\sigma_{xy}(3\omega_0)|}{|\sigma_{xy}(\omega_0)|}$ 
& $\displaystyle\frac{|\sigma_{xy}(5\omega_0)|}{|\sigma_{xy}(\omega_0)|}$ 
\\
\hline
      25 &      0.938   &    14\%  &   4\%    \\
     125 &      0.428   &    15\%  &   6\%    \\
     525 &      0.106   &    10\%  &   3\%    \\
     725 &      0.075   &    7\%   &   2\%    \\
     825 &      0.066   &    5\%   &  0.2\%   \\
\multicolumn{4}{c}{Case II}\\ 
\hline
$\hat y$ 
& $|\sigma_{xy}(\omega_0)|$ 
& $\displaystyle\frac{|\sigma_{xy}(3\omega_0)|}{|\sigma_{xy}(\omega_0)|}$ 
& $\displaystyle\frac{|\sigma_{xy}(5\omega_0)|}{|\sigma_{xy}(\omega_0)|}$ 
\\
\hline
   12.5 &      2.097   &    10\%   &    3\% \\
   62.5 &      0.594   &    13\%   &    5\% \\
   112.5&      0.294   &    13\%   &    3\% \\
   212.5&      0.139   &     9\%   &    1\% \\
   362.5&      0.078   &     3\%   &    1\% \\
\hline\hline
\end{tabular}
\end{table}
\begin{figure}[t]
\includegraphics[scale=1]{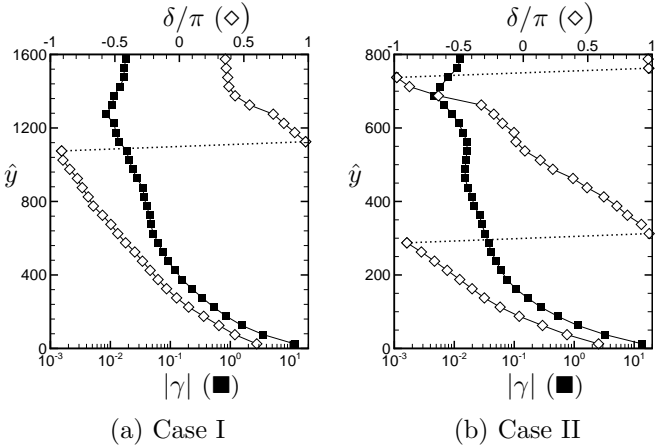}
\caption{
The spatial variations of $|\gamma|$ and $\delta$ in Eq. (\ref{eq_22}) for (a) Case I and (b) Case II.
The black square $\blacksquare$ shows $|\gamma|$ and the diamond $\Diamond$ shows $\delta/\pi$.
}\label{fig9}
\end{figure}
Figure \ref{fig9} shows the spatial variations of the amplitude of the local strain $|\gamma^l|$ and local phase retardation $\delta^l$ in Eq. (\ref{eq_22}) for Case I and Case II.
The profiles of the amplitude of the local strain $|\gamma|$ between the plates are similar in Case I and II.
These amplitude profiles increase rapidly in the boundary layer near the oscillating plate, and have values much greater than unity, $|\gamma|\sim 10$, in the vicinity of the plate.
Thus, the polymer chains are quite deformed near the oscillating plates, although the amplitude of strain to which the system is subjected via the oscillating plates is not so high; $\Gamma_0=0.5$ in both cases.

\begin{figure}[t]
\includegraphics[scale=1]{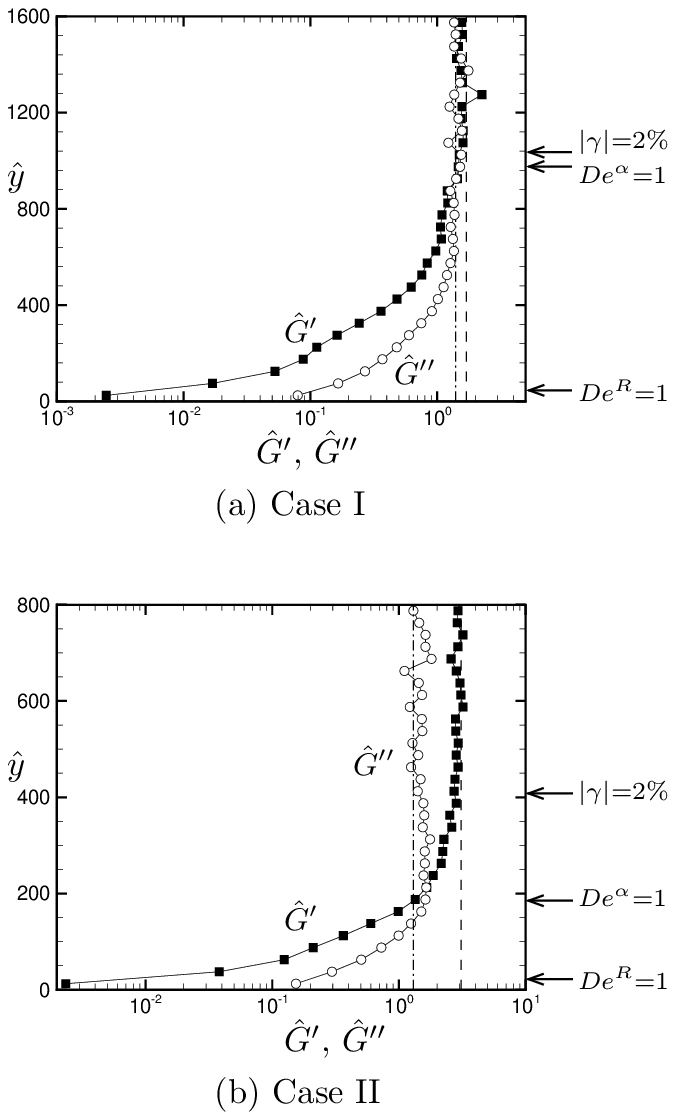}
\caption{
The spatial variations of the local moduli $\hat G'$ ($\blacksquare$) and $\hat G''$ ($\bigcirc$) for (a) Case I and (b) Case II.
The dashed and dash-dotted lines show the values of $\hat G'$ and $\hat G''$ for the linear regime 
($\hat G'=1.7$ and $\hat G''=1.4$ for Case I and $\hat G'=3.1$ and $\hat G''=1.3$ for Case II), respectively.
The linear moduli are calculated by the non-equilibrium MD simulations with small strains $0.005 < |\gamma| < 0.01$.
The left arrows on the right-side vertical axis show the positions where the local Deborah numbers, shown in Fig. \ref{fig10}, are equal to unity and
the position where the local strain is $|\gamma|=2\%$.
}\label{fig10}
\end{figure}
Figure \ref{fig10} shows the spatial variations of the local storage modulus and loss modulus for Case I and Case II.
Shear thinning is seen near the plate; both moduli $G'$ and $G''$ decrease near the oscillating plate.
In the close vicinity of the oscillating plate, the storage modulus $G'$ is much smaller than the loss modulus $G''$, $G' \ll G''$. 
Hence, the melt behaves as a viscous fluid. 
The storage modulus grows rapidly with the distance from the oscillating plate, and the visco-elastic crossover occurs at $\hat y\sim 1000$ for Case I and at $\hat y\sim 200$ for Case II. 
Both moduli attain their linear values, which are shown as dashed and dot-dashed lines in the figures, for a distance that is far from the oscillating plate where the local strains are less than about 2 $\%$ (See also Fig. \ref{fig9}). 
Thus, the local rheology of the melt can be divided into three regimes, i.e., the viscous fluid, visco-elastic liquid, and visco-elastic solid regimes. 

\begin{figure}[t]
\includegraphics[scale=1]{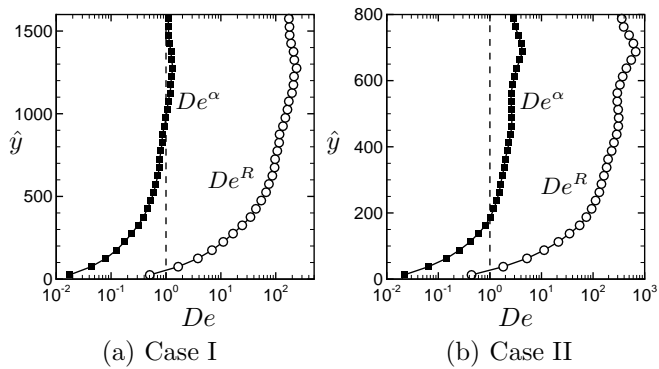}
\caption{
The spatial variations of the local Deborah numbers (defined via the Rouse relaxation time $\tau_R$ and the $\alpha$ relaxation time $\tau^\alpha$)
$De^R=\omega_0\tau_R$ ($\bigcirc$) and $De^\alpha=\omega_0\tau^\alpha$ ($\blacksquare$).
(a) The result for Case I and (b) that for Case II.
The dashed line represents $De=1$.
}\label{fig11}
\end{figure}
These regimes may be also characterized by the two ``local'' Deborah numbers. 
One is defined by the local Rouse relaxation time $\tau_R$ of the melt and the angular frequency of the plate $\omega_0$, $De^R$=$\omega_0\tau_R$, and the other is defined by the local $\alpha$ relaxation time $\tau_\alpha$ and the angular frequency $\omega_0$, $De^\alpha$=$\omega_0\tau_\alpha$. 
Note that the local Rouse and $\alpha$ relaxation times vary according to the local strain rate $\dot \gamma$, $\tau=\tau(\dot\gamma)$. 
Figure \ref{fig11} shows the spatial variation of the local Deborah number $De^R$ and $De^\alpha$, where the local relaxation times $\tau_R$ and $\tau_\alpha$ are calculated by substituting the values of $|\dot \gamma|^l$, which are obtained by Eq. (\ref{eq_21}) and equation below Eq. (\ref{eq_22}), into the fitting functions for the relaxation times for the simple shear flows obtained in Ref. \onlinecite{art:02YO}.
In Fig. \ref{fig10}, the positions at which the local Deborah numbers become equal to unity are shown by the left arrows. 
It is seen that the melt behaves as a viscous fluid, $G''\gg G'$, for $De^R\lesssim 1$, while the visco-elastic properties become pronounced for  $De^R\gtrsim 1$. 
This feature is also consistent with the rheology diagram for a model polymer melt obtained from Ref. \onlinecite{art:06VB}.
It is also seen that the crossovers from the liquid-like regime, $G''>G'$, to the solid-like regime, $G'>G''$, take place at $De^\alpha\sim 1$.

\section{Summary and outlook}
The behaviors of supercooled polymer melt in creep and recovery, pressure-driven flow and oscillating flows between the parallel plates are investigated numerically by using a hybrid simulation of MD and CFD. 
In the present hybrid method, the memories of molecular configurations of local fluid elements are traced at the microscopic level so that the visco-elastic motion of the melt is correctly reproduced.

The flow profiles of the melts are quite different from those of the Newtonian fluid.
In the creep simulation, we demonstrate the simple visco-elastic motion of the melt.
The nonlinear time evolution of strain of the system at the beginning of creep motion, such as the delayed elastic deformation, is reproduced.
After removing the stresses on the plates, the polymer melt recovers by an elastic strain stored in the creep motion. 
We also compare the result given by the present hybrid method with that given by the Jeffreys model for a constitutive relation.
It is seen that the result from the hybrid method is fitted well with that given by the Jeffreys model, although the fitting parameters in the Jeffreys model differ from each other in the creep and recovery motions since the mechanical property of the melt is quite sensitive to motion states.
(See Fig. \ref{fig2}.)
In pressure-driven flow, shear thinning occurs near the plates, where the local shear rates are much larger than those in the middle of the plates.
Shear thinning is enhanced as the pressure gradient increases or the temperature decreases, so that the velocity profile is increasingly flatter rather than parabolic at the middle between the plates.(See Figs. \ref{fig4} and \ref{fig5}.)
We also investigate the microscopic configurations of polymer chains at local points.
The polymer chains are quite elongated in the flow direction near the plate, while in the middle between the plates, configurations of polymer chains are rather incoherent. (See Fig. \ref{fig6}.)
For oscillating plates, we clarify that the viscosity of the melt becomes thin near the plates, and thus the boundary layer of the melt also becomes much thinner than that of the Newtonian fluid. (See Figs. \ref{fig7} and \ref{fig8}.)
The local rheological properties of the melt also vary considerably in the viscous boundary layer, so that three different regimes form between the oscillating plates, i.e., the viscous fluid, visco-elastic liquid, and visco-elastic solid regimes.
It is also found that, in the viscous fluid regime, the local Deborah number defined via the Rouse relaxation time and the angular frequency of the plate is approximately less than unity, $De^R\lesssim 1$. 
The crossover between the liquid-like and solid-like regimes takes place around the position where the local Deborah number (defined via the $\alpha$ relaxation time and the angular frequency) is approximately equal to unity, $De^\alpha\sim 1$. (See Figs. \ref{fig10} and \ref{fig11}.)

In the present hybrid method, the long-range correlations of the macroscopic quantities that are difficult to treat at the MD level are involved at the CFD level via the macroscopic momentum transport equation.
Thus, although each MD simulation is performed independently at each time step of CFD, the polymers in different MD cells are also correlated with each other via the macroscopic momentum transports.
The complicated mechanical properties depending on the microscopic configurations of polymer chains for which it is difficult to construct the model constitutive relation are also correctly involved in the CFD simulation, since the microscopic motions of polymers are resolved in the MD simulations associated with each mesh node of the CFD calculation according to the local flow field.
Hence, the present hybrid method is expected to be applicable both to problems for the large-scale flows which are out of range of the MD simulation, and the complex flows, which do not have any known model constitutive relations.

Compared with running a full MD simulation, the present hybrid simulation can save computation time as to the spatial domain, although it does not accelerate computation time with respect to the the temporal domain.
The efficiency of this hybrid simulation is represented by a saving factor defined by the ratio of the mesh size of the CFD simulation $\Delta x$ to the cell size of the MD simulation $l_{\rm MD}$, $\Delta x/l_{\rm MD}$.
In the present simulations, the saving factors are $\Delta x/l_{\rm MD}$=4.7 for the creep and recovery, $\Delta x/l_{\rm MD}$=5 for pressure-driven flow and Case I in the oscillation problem, and $\Delta x/l_{\rm MD}$=2.5 for Cases II--IV in the oscillation problem.
In addition to the saving factor $\Delta x/l_{\rm MD}$, the present hybrid method has also unique in that it is quite suitable as a parallel computational algorithm since the MD simulations associated with each mesh of the CFD, which represent a large part of the total simulation time, are performed independently.

\begin{figure}[t]
\includegraphics[scale=1]{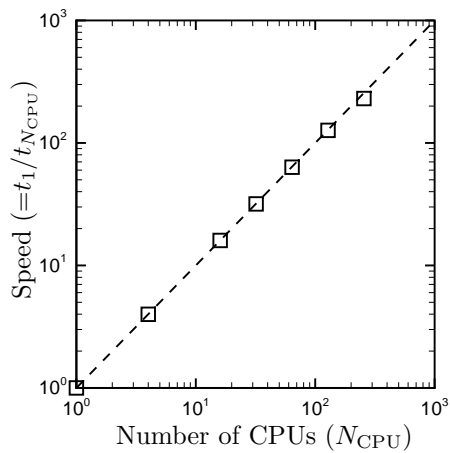}
\caption{
Benchmarks for parallel computations using the hybrid simulation method.
$N_{\rm CPU}$ is the number of CPUs and $t_{N_{\rm CPU}}$ the computational time of parallel computation with $N_{\rm CPU}$ CPUs.
The speed of parallel computation with $N_{\rm CPU}$ CPUs is measured by the ratio of the computational time using one CPU to that with $N_{\rm CPU}$ CPUs.
Parallel computations for $N_{\rm CPU}$=4, 16, 32, 64, 128, and 256 were performed on Fujitsu HX600 at T2K Open Supercomputer at Kyoto University.
}\label{fig12}
\end{figure}
We carry out benchmarks for parallel computations for the hybrid simulation method.
In the benchmarks, the creep simulations with $2\hat H=12800$ and $\hat \sigma_0$=0.1 are performed for 1,000 of CFD, $\hat t=[0,1000]$.
The distance between the plates $2H$ is divided into 257 slits.
Thus 256 MD cells with $\hat l_{\rm MD}=10$ are associated with each node of the slits.
The parallelization algorithm is applied to the process to calculate local stresses by MD simulation.
Hence, in the parallel computations with $N_{\rm CPU}$ CPUs, each CPU is assigned to the calculations of (256/$N_{\rm CPU}$) local stresses at each time step of the CFD.
Figure \ref{fig12} shows the result of the benchmarks.
The parallel computations are performed nearly ideal until the number of CPU $N_{\rm CPU}$is one-hundred twenty-eight, $N_{\rm CPU}=128$.
The efficiency of parallel computation $\varepsilon$ defined by $\varepsilon=(t_1/t_{N_{\rm CPU}})/N_{\rm CPU}$, where $t_{N_{\rm CPU}}$ is the computational time for $N_{\rm CPU}$ parallelization, is more than 99\% up to $N_{\rm CPU}=128$ and 90\% for $N_{\rm CPU}=256$.
Parallelization efficiency decreases a bit at $N_{\rm CPU}$=256 in the present benchmark since CFD calculation and communication between CPUs increases relative to the total computation.
However, for the simulations of large-scale and slow dynamics, we can improve the parallelization efficiency even for a large number of CPUs by setting the slit size and time-step size of the CFD calculation, $\Delta x$ and $\Delta t$, large while keeping the ratios $\Delta x/l_{\rm MD}$ and $\Delta t/t_{\rm MD}$ constant (because the ratio of CFD calculation computation time to the total time decreases.)
Although the total computation time dose indeed increases, the efficiency of the hybrid simulation compared to the full MD simulation does not change.
Thus we can expect to carry out quite high-performance parallel computations even for thousands or tens of thousands of CPUs.

In the present paper we are involved with only one-dimensional problems.
Future work includes the development of the hybrid method for two- or three-dimensional simulations.
In order to extend the present method to the two- or three-dimensional problems, we must treat the advection of memory of polymer chain configurations on the local fluid element.
For two-dimensional or three-dimensional flows of visco-elastic fluids, many novel simulation methods have been proposed.\cite{art:86LTI,art:86LTII, art:90RAB, art:98B,art:97LPO}
Incorporating the present hybrid technique into well-established simulation methods for the visco-elastic flows represents an important direction for future research.
The extension of this technique to the coupling of macroscopic heat and mass transfers is also important to the investigation of various physical problems involved in polymer melts.

\end{document}